# Direct Observation of Transverse and Vortex Metastable Magnetic Domains observed in Cylindrical Nanowires

C. Bran[1], J.A. Fernandez-Roldan[1], E.M. Palmero[1], E. Berganza[1], J. Guzman[1], R.P. del Real[1], A. Asenjo[1], A. Fraile Rodríguez[2], M. Foerster[3], L. Aballe[3], O. Chubykalo-Fesenko[1] and M. Vazquez[1]

[1]Institute of Materials Science of Madrid, CSIC, 28049 Madrid, Spain

[2]Departament de Física de la Matèria Condensada and Institut de Nanociència i Nanotecnologia (IN2UB), Universitat de Barcelona, 08028 Barcelona, Spain

[3] Alba Synchrotron Light Facility, CELLS, Barcelona, Spain

*We present experimental evidence of metastable magnetic domains in cylindrical CoNi nanowires. Transverse (perpendicular) domains have been previously observed only in flat nanostrips. Here they are found together with more conventional vortex domains along the same cylindrical nanowire, denoting a bistable system with similar free energies. The surface and the inner magnetization distribution in both types of domains are analyzed by Photo-Emission Electron Microscopy (PEEM) with X-ray magnetic circular dichroism (XMCD) contrast, and hysteresis loop in individual nanowires are measured by Magneto-optical Kerr effect. These experimental data are understood and compared with complementary micromagnetic simulations. The transverse domain states are potentially more important for sensing and recording applications than commonly observed vortex domains since they create larger surface stray fields.*

Magnetic domains and the magnetization reversal in ferromagnetic cylinders is a classical topic in nano-magnetism, where geometry by itself leads to unique consequences. The high aspect ratio defines a magnetic shape anisotropy which promotes longitudinal domains. On the other hand, the circular symmetry often enables less common domains where magnetic moments are not aligned parallel to each other but following a circumferential path. In addition, the axis of the cylinder introduces a mathematical singularity that is responsible for unusual Bloch-point domain walls and reversal mechanisms. All these peculiarities are even more pronounced in nanoscale wires. Here, we report on cylindrical CoNi nanowires with designed crystal symmetry and tailored magnetic anisotropy which show transverse magnetic domains through the whole wire cross section, including surface and inner regions, coexisting in the same wire with vortex domain segments. This fact is reproduced by complementary micromagnetic simulations.

Cylindrical magnetic nanowires, NWs, as single nanostructures or as part of 3D ordered architectures, are considered for applications in nanotechnology areas such as magnetic recording, microwave devices, functionalization for bio-magnetics and, more recently, for thermomagneto-electric devices[1-3]. They are proposed in spintronics as alternative to planar nanostrips owing to their specific advantages such as the possibility to tailor the shape of domain walls (DW) by adjusting the geometry[4], or their stability during motion[5] and the suppression of the Walker breakdown[6,7]. The diagram of magnetic states as a function of the NW aspect ratio has been considered from a micromagnetic viewpoint which essentially consists of axial, transverse and vortex states[8,9] with transverse (in-plane) states occurring in ultra-thin dots only. The transverse domains are frequently observed in magnetic nanostrips[10]. However, transverse domains in cylindrical nanowires have not been reported so far not even from micromagnetic point of view.

The synthesis of cylindrical NWs inside porous templates with engineered composition and geometry is achieved by an in-expensive electrochemical route[11]. Complex local magnetic configurations have previously been imaged at the surface cap ends of nanowires inside the template by magnetic force microscopy in FeCo nanowires[12] and by electron holography in Co-based[13] nanowires. At the wire ends, magnetization locally redistributes in order to reduce the stray field energy.

All the mentioned technological applications rely on the understanding and control of the magnetic configurations of the individual magnetic units, determined by the composition through the balance between the geometrical shape, magneto-crystalline and magneto-elastic anisotropies. Much less attention has been paid to individual nanowires owing to difficulties to release them properly from the template and to perform reliable experimental work, usually in expensive and/or unavailable techniques, especially considering the additional difficulties introduced by the circular cross section. Only very recently an interest is growing to search for the deeper understanding of the domain structure and domain wall motion in cylindrical nanowires. Experimental imaging of the surface magnetic state has been carried out at remanence or under *in-situ* applied magnetic field by magnetic force microscopy (MFM)[14] and magneto-optical Kerr effect (MOKE)[15]. In addition, electron holography (EH) supplies information of the magnetic flux distribution of the internal magnetic structure apart from the stray fields outside the nanowires[16,17]. These techniques have been specifically used to investigate the magnetization reversal mechanism by domain wall nucleation and propagation in Fe, Co, and Ni based nanowires (single element, alloys or multisegmented) with various modulations in diameter or in composition. In all these cases, the pinning of the domain wall motion is attempted by local transition regions where diameter or composition changes.

Photo Emission Electron Microscopy combined with X-ray Magnetic Circular Dichroism[18] (XMCD-PEEM) has proved recently to be a unique technique to explore the magnetization configuration of individual cylindrical nanowires. Thanks to the partial transmission of a grazing incidence X-ray beam through the wire, the core magnetic state is mapped onto the substrate, providing simultaneous information of the magnetization distribution at the surface (direct photoemission from the wire) and inside the nanowires (photoemission from the substrate). XMCD-PEEM has been successfully employed to identify Bloch-point domain walls in Permalloy nanowires characterized by axial domains[19]. Vortex domains with circumferential magnetization at the surface and longitudinal at the core have been very recently confirmed in Co nanowires[20].

In this work we synthesized CoNi cylindrical nanowires with different Co content and designed magnetocrystalline anisotropy. By analogy to nanostrips where perpendicular magnetic anisotropy is responsible for transverse domains, we are able to design such domains in cylindrical NWs. Such magnetic domain structure could offer an alternative for high-density and three-dimensional data storage devices in advanced magnetic memories since it is preferable to more common vortex domains which almost do not create surface charges. Our nanowires are 120 nm in diameter and several micrometers in length. In alloys with less than 50% Co (e.g., $Co_{35}Ni_{65}$ nanowires), a *fcc* phase is detected and the predominant shape anisotropy results in a single axial domain. For intermediate composition (e.g., $Co_{65}Ni_{35}$) both *fcc* and *hcp* phases are identified[21], while for Co-rich nanowires (e.g., $Co_{85}Ni_{15}$) only *hcp* phase is detected. In all cases the energetic balance between shape and crystalline anisotropy determines the presence of transverse and vortex hybrid domain structures. Additional information on the synthesis and general characterization of nanowires is given in the *Supplementary* information.

Longitudinal MOKE hysteresis loops of individual $Co_{85}Ni_{15}$ and $Co_{65}Ni_{35}$ nanowires, measured in NanoMOKE™ 2 setup, and shown in Figure 1(a) were recorded under applied field parallel to the nanowire axis. For $Co_{85}Ni_{15}$ nanowire only a minor loop with reduced hysteresis and longitudinal susceptibility has been measured due to limitations of the maximum applied magnetic field in our setup. This magnetic behavior is consistent with the presence of an *hcp* crystalline phase with strong magnetocrystalline anisotropy value causing the reorientation of the magnetization easy axis towards a nearly perpendicular orientation. In the case of $Co_{65}Ni_{35,}$ a larger hysteresis is observed together with significant higher remanence. Figure 1(b) shows the simulated hysteresis loops for both $Co_{85}Ni_{15}$ and $Co_{65}Ni_{35}$ nanowires. Longitudinal MOKE hysteresis loops provide us with first information on individual nanowires although data should be carefully considered as they are taken under maximum field of 0.50 kOe, not large enough to saturate magnetically the samples. Consequently, a comparison between experimental data and simulations cannot be made straightforward.

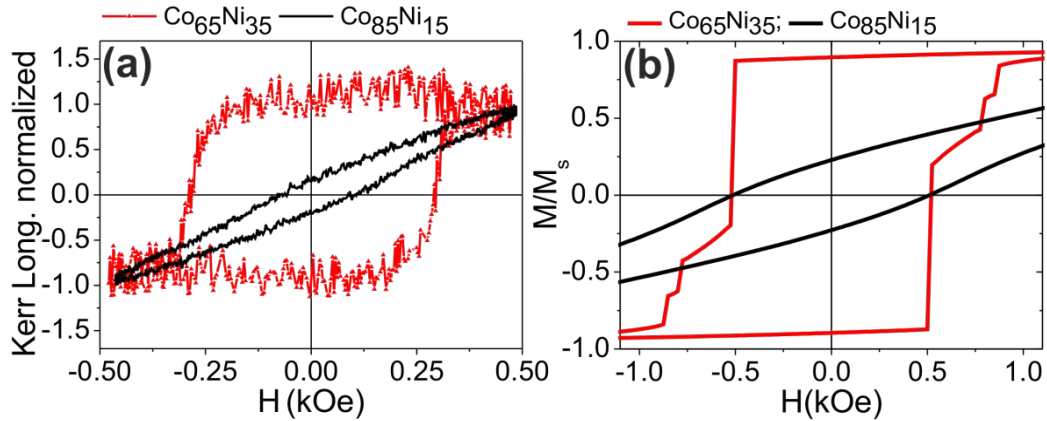

*Figure 1. a) MOKE hysteresis loops of $Co_{85}Ni_{15}$ (black) and $Co_{65}Ni_{35}$ (red) nanowires under magnetic field applied parallel to the nanowire axis; b) Simulated hysteresis cycles of $Co_{85}Ni_{15}$ (black) and $Co_{65}Ni_{35}$ (red) nanowires.*

XMCD-PEEM experiments were performed at the CIRCE beamline of the ALBA Synchrotron[22]. For Ni-rich nanowires ($Co_{35}Ni_{65}$), a highly uniform axially magnetized state is obtained at remanence (see Fig. S2, *Supplementary Information*). Here, we focus on Co-rich nanowires where the images reveal the most interesting and complex magnetic domain structure. Figure 2 shows XMCD-PEEM images of individual $Co_{85}Ni_{15}$ and $Co_{65}Ni_{35}$ nanowires measured in a nearly perpendicular configuration of the X-ray propagation vector with respect to the nanowire axis. The marked regions of the images labeled as "wire" and "shadow" corresponds respectively to photoemission from the nanowire surface and that from the substrate after transmission through the wire core (see Fig. S3, *Supplementary Information*, for a schematic view of the contrast modes of surface/core).

In Figure 2(a), we observe two different magnetic contrasts along a $Co_{85}Ni_{15}$ nanowire. At the surface, it consists of alternating segments with opposite bright/dark contrast along the nanowire. Such image indicates that the magnetic moments are oriented parallel and anti-parallel to the X-ray propagation vector, respectively, that is, perpendicular to the nanowire axis. The shadow contains segments with an opposite contrast to that of the "wire" (surface). The narrow strip of alternating bright/dark contrast observed at the upper edge, with opposite contrast and same periodicity as that of the image obtained in direct photoemission corresponds to transmitted photons through the top surface of the nanowire. Notice that for the same magnetic moment orientation, opposite contrast is expected between direct and transmitted signal (See Fig. S3, Supplementary information). From the bright/dark pattern of the whole shadow, the presence of a vortex domains structure with alternating chirality along the nanowire surface is concluded, similar to those observed in Co nanowires[20]. Note that the length of such vortex domains is not constant along the nanowire.

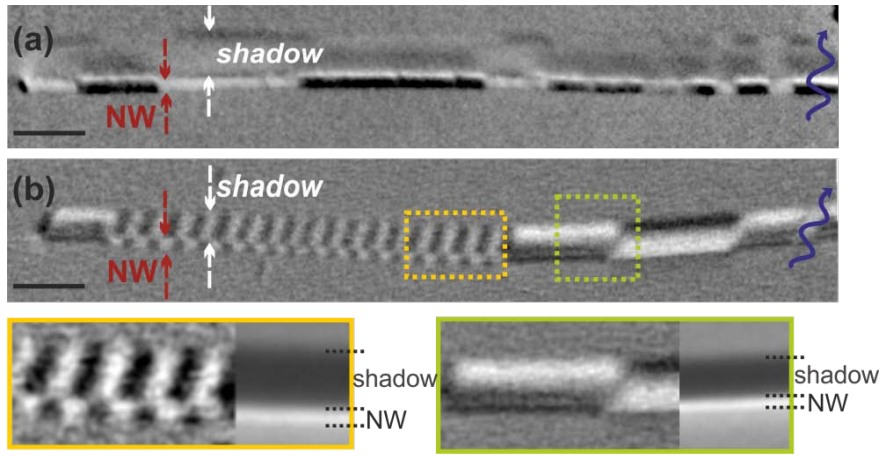

*Figure 2. XMCD-PEEM images of $Co_{85}Ni_{15}$ (a) and $Co_{65}Ni_{35}$ (b) nanowires. The scale bar is 1 µm, and the arrows (dark blue) indicate the incident X-ray beam. Inset figures at the bottom are enlarged images of the transverse and vortex domains in the corresponding regions of $Co_{65}Ni_{35}$. Right-side in the insets: Direct (XAS) image of the same wire.*

However, the most striking configurations are observed for a $Co_{65}Ni_{35}$ nanowire where the magnetic contrast at the surface shows two different regions. This is depicted in Figure 2(b), where the X-ray absorption (XAS) images are used as a reference to correlate the position of the wire and the shadow.

i) On the right side of the wire, a sequence of segments with opposite contrast similar to those in Figure 2(a) is observed. The "shadow" consists of segments with an opposite contrast to that of the "wire" (surface) and increasing intensity as the edge of the wire is reached (see the inset) indicating that the circular magnetization pattern penetrates nearly fully into the nanowire. Thus, we conclude the presence of vortex domains with alternating chirality. Again, their length is not constant along the nanowire.

ii) On the left side of the wire, a regular sequence of segments with much shorter periodicity and alternating contrast (i.e., bright/dark at the surface, dark/bright in the core) is observed (see also inset in Fig. 2b). The contrast in the shadow is opposite to that at the surface, and remains constant in the transversal direction (see the inset), showing clearly that the magnetization state is homogenous along the complete circular cross section of the wire. This is clearly different from the typical vortex domains in Figure 2(a), and it is interpreted to correspond to transverse domains with significant component of magnetization in the perpendicular direction to the nanowire axis and alternating orientation. The width of the observed transverse domains is estimated to be about 150 nm and very regular. Both can be understood to be a result of the large local stray field associated to a transverse domain which is compensated by alternating domains. Note that in contrast, a vortex domain represents a flux closure state with little or no local stray field.

This hybrid domain structure, i.e. the coexistence of transverse and vortex domains in the same wire, anticipates that the local domain structure is not uniquely defined by the composition, but is determined by a very subtle energy differences, defining its metastable character. For a deeper understanding, micro-magnetic simulations of cylindrical nanowires were performed using the mumax3 code[23] for $Co_xNi_{1-x}$ alloy nanowires (x=85, 65 and 35). Based on our experimental data, a polycrystalline *fcc* (111) cubic symmetry for $Co_{35}Ni_{65}$, and a single *hcp* crystal with easy axis nearly perpendicular to the NW axis for $Co_{85}Ni_{15}$ nanowires have been considered. For intermediate $Co_{65}Ni_{35}$ composition, we took an *hcp* polycrystalline structure with "*c*" axis at 65º with respect to the NW axis and randomly distributed azimuthal angle. The equilibrium state is obtained by relaxation from a starting random magnetic configuration as corresponding to an as-prepared state. Further details are given in the *Supplementary Information*.

The simulation results corresponding to one of the possible configurations for the $Co_{85}Ni_{15}$ nanowire is shown in Figure 3(a) where the equilibrium state is formed by a non-systematic distribution of vortex and transverse domains along the nanowire. The pattern consists of vortex domains with opposite chiralities and different lengths between shorter transverse domains with opposite directions close to the ends of the nanowire. The vortex structures extend over 68% of the nanowire. In short, the major section of the wire is formed by vortex domain structure with random size and chirality distribution, similarly to the XMCD-PEEM image shown in Figure 2(a).

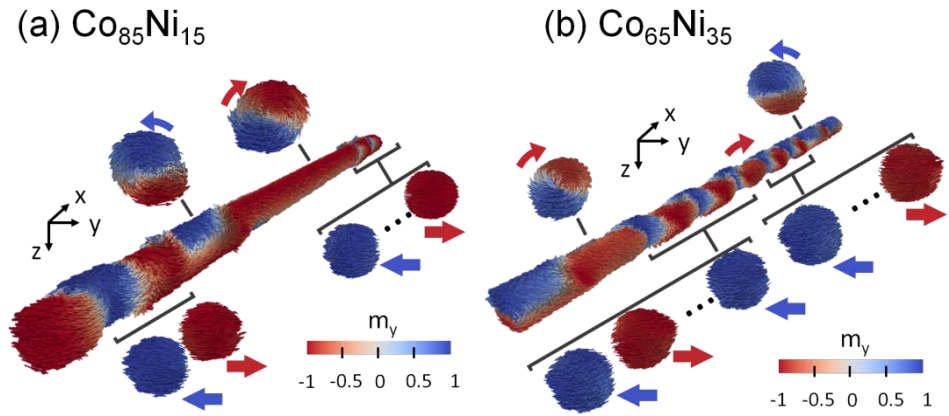

*Figure 3. Modeled magnetization configuration for $Co_{85}Ni_{15}$ (a) and $Co_{65}Ni_{35}$ (b) nanowires. Insets present the cross sections at the marked sites. Color scale shows the transverse component of magnetization.*

In the case of $Co_{65}Ni_{35}$ nanowire (Figure 3(b)) a more complex domain structure is obtained. Again, a hybrid magnetic structure of transverse and vortex domains is observed along the nanowire. However, now the transverse domains occupy a larger fractional volume, and the vortex domains mostly appear close to the ends. Around 52% of the nanowire length is filled by segments with opposite alternating transverse domains regularly distributed with an estimated length over 90 nm, notably close to the experimental data.

Overall, simulations reproduce remarkably well the experimental XMCD-PEEM data although some points are not fully reflected. In $Co_{85}Ni_{15}$ nanowires, the simulated transverse domains were not observed experimentally. In $Co_{65}Ni_{35}$ nanowires, the hybrid system of transverse and vortex domains is well accounted for in the simulations, included the regular distribution transverse domains although their length, 90 nm in the simulation, is shorter than found in the experiment, 150 nm. We also note that in the Figure 2(b), the large zone with transverse domains changes into vortex domain close to the left end of the wire like in simulations where vortices are mainly found at the end.

Vortex domain structures in *hcp* Co nanowires with almost perpendicular magnetocrystalline anisotropy have been recently reported to minimize the total magnetic energy[20]. Present simulations additionally predict the coexistence of vortices and transverse domains in alloy nanowires by tuning the magnetocrystalline anisotropy with the Co content of the alloy in the Co-rich case. That has been fully experimentally confirmed in the case of $Co_{65}Ni_{35}$ nanowires. The coexistence of the two kinds of domains actually reflects the metastable character of nanowires with multiple states of similar energies. For the simulations, fine-tuning of saturation magnetization value, exchange stiffness constant and magnetocrystalline anisotropy constant have been considered to reach the optimal fitting to experimental XMCD-PEEM data and nanoMoke loop.

The vortex structure essentially minimizes the magnetostatic energy while the transverse one minimizes the exchange plus the anisotropy ones. The magnetostatic energy is additionally minimized by creation of multiple domains along the nanowire. Since the $Co_{85}Ni_{15}$ has larger saturation magnetization (shape anisotropy energy density, $K_{mag} \sim 510$ kJ/m$^3$) than $Co_{65}Ni_{35}$ ($K_{mag} \sim 360$ kJ/m$^3$), the formation of vortex domains is more favorable in $Co_{85}Ni_{15}$. However, one should also take into account that the anisotropy energy is partly minimized for the vortex domains by slight displacement of the core from the axial position[20]. Additionally, the domain walls between the vortex domains are more complex and require more exchange energy that those between the transverse domains. Consequently, the transverse domains become more favorable for a smaller value of the saturation magnetization. Since the magnetostatic energy is almost completely minimized by vortex domains, no addition minimization is needed and the domains are much larger than in the transverse case.

In conclusion, metastable hybrid states of transverse and vortex domains are confirmed to coexist in the same CoNi nanowire. The transverse magnetic domains with homogeneous magnetization perpendicular to the nanowire axis have been imaged by XMCD-PEEM. The data allows us to interpret the magnetization distribution at the surface and the inner region of the nanowires simultaneously. The coexistence of the metastable hybrid state is confirmed by micromagnetic simulations. The presence of the two types of domains in the same nanowire is understood considering the energy balance between transverse and vortex domain structures.

The design of transverse/vortex domains in cylindrical nanowires through suitable engineering of the magnetic anisotropy by small compositional changes opens an inexpensive route for high-density magnetic memories. The transverse domain states offer a novel alternatively for advanced magnetic sensing and recording with more potential than commonly observed vortex domains. First, the local stray field produced by transverse domains is stronger as compared to the vortex ones, which is important for reading the output signal. Secondly, due to the minimization of the magnetostatic energy the transverse domains have reduced wave length, consequently enabling a higher information density.

AUTHOR INFORMATION

**Corresponding Author**

*Cristina Bran, email: cristina.bran@icmm.csic.es

**Present Addresses**

E.M. Palmero: IMDEA Nanociencia, C/ Faraday, 9, 28049 Madrid, Spain

**Acknowledgements**

The study has been performed under the framework of the projects MAT2013-48054-C2-1-R and MAT2016-76824-C3-1-R supported by the MINECO from Spain. J.A.F-R acknowledges financial support from MINECO and the ESF through the "Ayudas para contratos predoctorales para la formación de doctores 2014". The PEEM experiments were performed at the CIRCE beamline of the ALBA Synchrotron with the collaboration of ALBA staff.

*Supplementary Information*

# Direct Observation of Transverse and Vortex Metastable Magnetic Domains observed in Cylindrical Nanowires

C. Bran[1], J.A. Fernandez-Roldan[1], E.M. Palmero[1], E. Berganza[1], J. Guzman[1], R.P. del Real[1], A. Asenjo[1], A. Fraile Rodríguez[2], M. Foerster[3], L. Aballe[3] O. Chubykalo-Fesenko[1] and M. Vazquez[1]

[1] Institute of Materials Science of Madrid, CSIC, 28049 Madrid, Spain

[2] Departament de Física de la Matèria Condensada and Institut de Nanociència i Nanotecnologia (IN2UB), Universitat de Barcelona, 08028 Barcelona, Spain

[3] Alba Synchrotron Light Facility, CELLS, Barcelona, Spain

*Controlled synthesis and structural characterization*

CoNi nanowires (NWs) were prepared by filling self-assembled pores of anodic aluminum oxide (AAO) templates by electroplating. The templates were obtained by hard anodization in oxalic aqueous solution (0.3M) containing 5 vol.% ethanol at a constant temperature of 0 ºC. During the anodization, a constant voltage of 80 V was firstly applied for 900 s to produce a protective aluminum oxide layer at the surface of the disc, which avoids breaking or burning effects during the subsequent hard anodization [1]. After that, the voltage was steadily increased (0.08 V/s) up to 130V and kept constant for 3600 s. Nanopores with 120 nm in diameter and 60 μm in length were thus obtained. Afterwards, the residual Al and the alumina barrier layer at the bottom of the foils were chemically etched, and an Au layer was sputtered to serve later as an electrode for final electroplating of NWs. $Co_xNi_{100-x}$ (35≤x≤85) alloy NWs were grown into the nanopores of AAO templates, at room-temperature, by DC electrodeposition using the electrolytes: 0.12-0.09M $CoSO_4$ + 0.063-0.084M $CoCl_2$ + 0.057-0.095M $NiSO_4$ + 0.058-0.084M $NiCl_2$ + 0.32M $H_3BO_3$ [2]. The pH value was kept constant at about 3.0. For MOKE, MFM, and XMCD-PEEM measurements the individual NWs were released from the membranes by dissolving the alumina using an aqueous solution composed by chromic oxide and phosphoric acid.

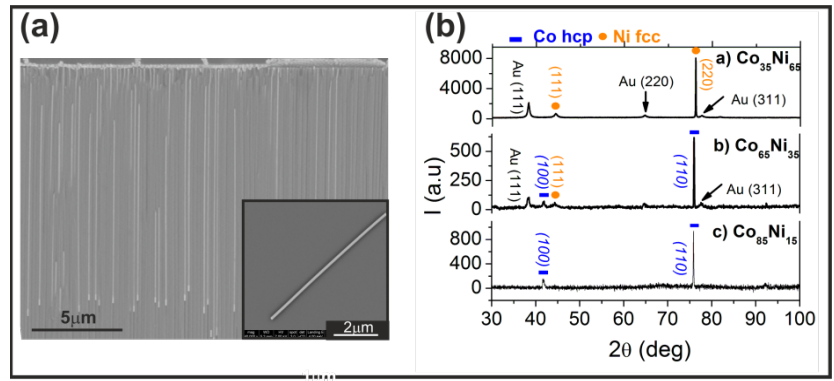

Figure S1. a) Cross-sectional SEM image of a CoNi NW array. The inset shows an individual NW on a Si substrate; b) XRD spectra of arrays of $Co_xNi_{100-x}$ NWs;

From Scanning Electron Microscopy (SEM) (see Fig. S1a) measurements we determined the NW diameters to be about 120 nm and the respective lengths: L=8 μm ($Co_{35}Ni_{65}$), L=18 μm ($Co_{65}Ni_{35}$) and L=18 μm ($Co_{85}Ni_{15}$). The composition of NW alloys, $Co_{35}Ni_{65}$, $Co_{65}Ni_{35}$ and $Co_{85}Ni_{15}$, was determined by Energy-Dispersive X-ray spectroscopy (EDX). The structural characterization was performed by X-ray diffraction (XRD) using a PANalytical X'pert Pro X-ray diffractometer in Bragg-Brentano geometry.

The XRD patterns of AAO membranes filled with $Co_xNi_{100-x}$ NWs are presented in Figure S1b. In $Co_{35}Ni_{65}$ NWs, Ni *fcc* (220) and (111) peaks are present. Additional peaks at 38.1, 64.5 and 77.5 deg. belong to *fcc* Au sputtered on the back side of the sample. For $Co_{65}Ni_{35}$, the *hcp* phase with (110) peak at 75.8 deg. is detected as well as a smaller peak ascribed to (220) *fcc* Ni. In the case of $Co_{85}Ni_{15}$, only the *hcp* phase with (110) and (100) peaks is observed. A transition from *fcc* cubic structure for Ni-rich NWs towards *hcp* hexagonal phase for Co-rich alloy is thus concluded.

*Magnetic characterization of individual nanowires*

For magnetic force microscopy (MFM) imaging, a system from Nanotec Electronica was used, working with amplitude modulation control mode and the two passes method. The phase-locked loop was enabled to track the resonance frequency of the oscillating cantilever and hence, the frequency shift signal is recorded to obtain the MFM images. The measurements were performed with microchips Multi75M-G, from Budget Sensors (resonance frequency 75 kHz and force constant 3 N/m). The oscillation amplitude of the cantilever was set in 10-15 nm and a typical retrace distance of 40 nm was chosen. Figure S2a presents the MFM image of a $Co_{35}Ni_{65}$ NW where a uniform contrast is observed along the nanowire length, indicating a nearly single longitudinal domain. The concentration of magnetic charges of opposite sign at the ends is ascribed to the so called dipolar contrast.

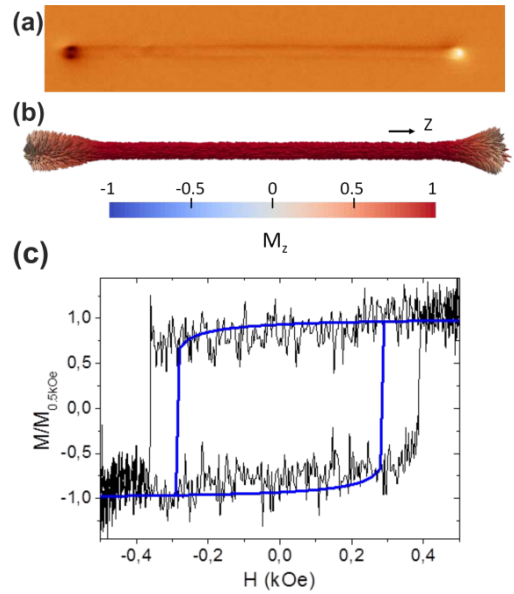

Figure S2. $Co_{35}Ni_{65}$ nanowire: (a) MFM image, (b) simulated magnetization distribution; and (c) MOKE (black) and simulated (blue) hysteresis loops.

MOKE hysteresis loops representing the normalized Longitudinal Kerr signal for a maximum applied field of 0.5 kOe were obtained using a Kerr effect magnetometer (NanoMOKE™ 2) with a nominal laser spot of 3 μm. The measurements were performed on different NWs from the same batch to confirm the representative behavior of depicted MOKE hysteresis loop (each loop was the result of 500 averaged cycles). Figure S2c shows the longitudinal hysteresis loop of a $Co_{35}Ni_{65}$ NW with bistable behavior between the two opposite magnetization directions in agreement with the single axial domain imaged in Figure S2a.

The XMCD-PEEM measurements were performed at the CIRCE beamline of the ALBA Synchrotron Facility (Barcelona, Spain). The samples are illuminated with circularly polarized X-rays at a grazing angle of 16° with respect to the surface, at the resonant $L_3$ absorption edge of Co (778 eV). The emitted photoelectrons used to form the surface image are proportional to the X-ray absorption coefficient and thus the element-specific magnetic domain configuration is given by the pixel-wise asymmetry of two PEEM images sequentially recorded with left- and right-handed circular polarization. Actually, the projection of the local magnetization on the photon propagation vector is imaged so that, in the XMCD image ferromagnetic domains with magnetic moments parallel or antiparallel to the X-ray polarization vector appear bright or dark, respectively while domains with magnetic moments at a different angle have an intermediate gray contrast. Furthermore, the particular cylindrical shape of the wires allows for a fractional amount of X-rays to be transmitted through the NW, generating photoemission from the Si substrate. By analyzing the circular dichroic or pseudo-magnetic contrast formed in transmission in the shadow area, information about the magnetic configuration in the bulk of the wire can then be obtained (for further details, see Refs. 16 and 20 in the main manuscript). For a better understanding of the XMCD-PEEM images, Figure S3 shows a schematic view of the contrasts from the NW and from the shadow. Notice that the dark contrast in transmission is equivalent to the bright one in direct photoemission, since the absorbed and the transmitted X-rays are complementary.

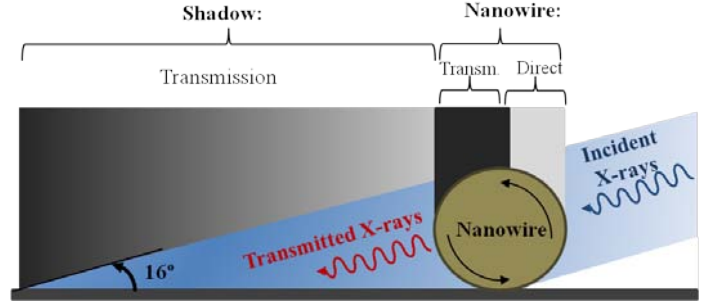

Figure S3. Schematic view for the interpretation of the contrasts observed in direct photoemission and transmission, dual in the case of a NW with circular spin configuration.

*Micromagnetic simulations*

Micromagnetic modeling in cylindrical NWs was performed using mumax3 code [3] for $Co_xNi_{100-x}$ alloys, with x = 85, 65 and 35. The NWs in the simulations were 1.5 µm long and 100 nm in diameter. A linear relation with composition was assumed for the saturation magnetization with $\mu_oM_{S,Co}$ = 1.8 T and $\mu_oM_{S,Ni}$ = 0.61 T for pure Co and Ni respectively, ($\mu_oM_{S,Co85}$ = 1.60, $\mu_oM_{S,Co65}$ = 1.35 T and $\mu_oM_{S,Co35}$ = 1.01 T), while the exchange stiffness constant was set to 26, 15 and 10 pJ/m, respectively [4-7]. For $Co_{35}Ni_{65}$ a *fcc* (111) cubic structure is assumed with a magnetocrystalline anisotropy constant, K= 2 kJm$^{-3}$ (much smaller than the estimated shape anisotropy constant, $K_{sh}$= 200 kJm$^{-3}$). Figure S2b shows the simulated remanence state for the $Co_{35}Ni_{65}$ nanowire, where a single axial domain is observed except for vortex-like structures at the ends (the simulated hysteresis loop together with the MOKE experimental one are shown in Fig. S2c). For $Co_{85}Ni_{15}$ nanowire, we assume an *hcp* hexagonal single crystal with c-axis at 88 deg. with respect to the NW axis and magnetocrystalline anisotropy constant $K_1$= 350 kJ/m$^3$ (shape anisotropy $K_{sh}$= 510 kJm$^{-3}$). An *hcp* polycrystalline structure was considered with *c*-axis at 65 deg. for $Co_{65}Ni_{35}$ with magnetocrystalline anisotropy constant $K_1$= 260 kJ/m$^3$, and shape anisotropy constant $K_{sh}$= 360 kJm$^{-3}$. For the proper analysis of experimental data, a fine tuning of magnetocrystalline anisotropy constant between 200 and 300 kJ/m$^3$ and *c*-axis up to 75 deg. was made for the intermediate $Co_{65}Ni_{35}$ alloy NW and exchange stiffness between 10 and 20 pJ/m. As-prepared remanent states were calculated by energy minimization starting at arbitrary initial configurations under no applied field. Finally, simulations indicate the different type of domain walls in vortex and transverse domain structures. 3D walls appear between vortex domains, and if the domain cores have opposite magnetization direction the wall goes through a Bloch point. In the transverse domain structure, domain walls are similar to the case of nanostripes, and they can be of Nèel or surface vortex type.